\documentclass{sig-alternate-10pt}
\usepackage{booktabs} 
\usepackage{epstopdf}
\usepackage{url}
\usepackage{subcaption}
\usepackage{graphicx} 
\usepackage[normalem]{ulem}
\usepackage{vmargin}
\usepackage{xcolor}

\useunder{\uline}{\ul}{}
\begin{document}
\title{The Internet Pendulum: On the Periodicity of Internet Topology Measurements}
\numberofauthors{3}
\author{
\alignauthor
Mattia Iodice\\
       \affaddr{Roma Tre University}\\
       \affaddr{Rome, Italy}\\
       \email{mattia.iodice3@gmail.com}
\alignauthor
Massimo Candela\\
       \affaddr{RIPE NCC}\\
       \affaddr{Amsterdam, The Netherlands}\\
       \email{mcandela@ripe.net}
\alignauthor
Giuseppe Di Battista\\
       \affaddr{Roma Tre University}\\
       \affaddr{Rome, Italy}\\
       \email{giuseppe.dibattista@uniroma3.it}
}
\maketitle

\begin{abstract}
Public databases of large-scale topology measures (e.g.\ RIPE Atlas) are very popular both in the research and in the practitioners communities. They are used, at least, for understanding the state of the Internet in real time, for outage detection, and to get a broad baseline view of the Internet evolution over time. However, despite the large amount of investigations, the dynamic aspects of these measures have not been fully understood. 
As an example, looking at time-series of such measures it happens to observe patterns that repeat at regular intervals.
More specifically, looking at a time-series of traceroutes involving certain source-target pairs it happens to observe that the paths follow alternations that repeat several times.
Have they the features of periodicity? What are their main characteristics?
In this paper we study the problem of detecting and characterizing periodicities in Internet topology measures. For this purpose we devise an algorithm based on autocorrelation and string matching. First, we validate the effectiveness of our algorithm in-vitro, on randomly generated measures containing artificial periodicities. Second, we exploit the algorithm to verify how frequently traceroute sequences extracted from popular databases of topology measures exhibit a periodic behavior. We show that a surprisingly high percentage of measures present one or more periodicities. This happens both with traceroutes performed at different frequencies and with different types of traceroutes. Third, we apply our algorithm to databases of BGP updates, a context where periodicities are even more unexpected than the one of traceroutes. Also in this case our algorithm is able to spot periodicities. We argue that some of them are related to oscillations of the BGP control plane.
\end{abstract}

\section{Introduction}

The availability of public databases of large-scale Internet topology measures (e.g.\ the traceroutes performed by RIPE Atlas~\cite{atlas} or by Caida Ark ~\cite{caida}) has deeply changed our possibility of understanding the state of the Internet in real time, our outage detection systems, and our methods for getting a broad baseline view of the Internet evolution over time. However, even if several papers have been written where such public databases are the leading actors, the dynamics of these topology measures have not been fully understood. This is important for many research topics and, because of the presence of measure changes that do not have anything to do with faults, is especially crucial for outage detection systems.

\begin{figure*}[tb]
  \includegraphics[width=\textwidth,height=4cm]{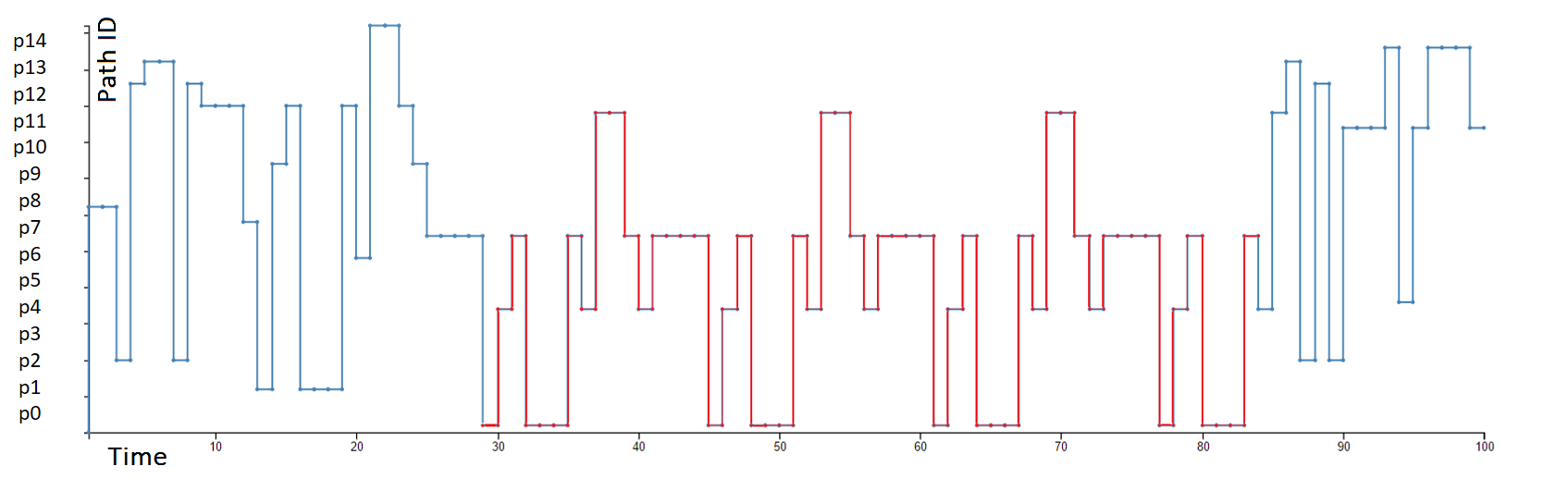}
  \caption{A time-series of paths originated by a sequence of traceroutes. Each path is assigned a Path-id in $\left[ p_0, p_{14} \right]$. The mapping between paths and IPv4 addresses is omitted for brevity. The paths have been recorded by RIPE Atlas on May 1st 2017 from 00:00 to 04:00 am. The probe identifier is 10039 and the target identifier is nl-ams-as43996.}
\label{fig:first-example}
\end{figure*}

When looking at time-series of traceroute paths (where a path is a sequence of IP addresses possibly containing asterisks) with one of the many popular visualization tools (see e.g.~\cite{cdds-dtvmal-13}) it happens to observe paths that follow alternations that repeat several times. Have they the features of the ``periodicity''? What are their main characteristics? As an example, the diagram of Fig.~\ref{fig:first-example} shows a periodic pattern. It represents a time-series of paths originated by a sequence of traceroutes performed between a specific probe-target pair. The presence of a periodicity, involving paths $p_0$, $p_4$, $p_7$, and $p_{11}$, is quite evident even if mixed with some ``noise.''

In the networking field we can have different types of periodicity. As an example we can have that the data plane of a router changes periodically, because of a control plane misconfiguration or because of some traffic-engineering requirement. Observe that we can have a data plane that has a periodic behavior that, because of an inauspicious timing is not revealed by topology measures. On the contrary we have topology measures that exhibit periodicity that do not correspond to any periodicity of the data plane. As an example, consider a sequence of Paris traceroutes~\cite{augustin2006avoiding} against an hash-based load balancer.

In this paper we study periodicities in Internet topology measures. We do not deal with the causes of such periodicites but we study their presence in public databases. First, we propose an algorithm, based on autocorrelation, string matching, and clustering, for detecting and characterizing periodicities. Second, we check the algorithm in-vitro, against a set of randomly generated time-series of traceroute paths. We show that the algorithm is very effective even in the presence of a certain amount of noise. Third, we exploit the algorithm to verify how frequently traceroute sequences extracted from popular databases of topology measures exhibit a periodic behavior. We show that a surprisingly high percentage of measures present one or more periodicities. This happens both with Paris traceroute and with traditional traceroute. This also happens with traceroutes performed at different frequencies. Finally, we slightly modify our algorithm and apply it to sequences of Border Gateway Protocol (BGP) updates. We concentrate on the prefixes that are most active in terms of recorded BGP updates and observe several periodicities in data extracted from the RIS service of RIPE. This is even more surprising than the results obtained by examining traceroutes. In fact, before being recorded by RIS \emph{collector peers}, the BGP updates traverse so many BGP routers, each with its own timing features, to make the synchronization required to spot periodicities unlikely. Among the periodicities we were able to distinguish some cases that are related to BGP control plane oscillations. As far as we know, this is the first time that this type of oscillation is detected in the wild.

The paper is organized as follows. In Section~\ref{sec:terminology-state-of-the-art} we introduce basic terminology and discuss the state-of-the-art. In Section~\ref{sec:data-set} we describe the data set of traceroutes used in the experiments. In Section~\ref{sec:methodology} we present our methodology. In Section~\ref{sec:vitro} we show the experiment conducted in-vitro, in Section~\ref{sec:traceroute-periodicity} we show the experiments on traceroutes, and in Section~\ref{sec:bgp-experiments} we show the experiments on BGP updates. Conclusions are in Section~\ref{sec:conclusions}.

\section{Basic Terminology and Related Work}\label{sec:terminology-state-of-the-art}

A \emph{periodic function} is a function that repeats its values in regular intervals. Function $f$ shows a periodic behavior~\cite{wiki:Periodic_function} with \emph{period} $P$ in an interval $T$ if $f(x+P)=f(x)$ for each $x \in T$. The values of $f$ in $T$ are the \emph{periodic pattern} of the periodicity. If there exists a least positive constant $P$ with this property, it is called the \emph{fundamental period}. Notice that a periodic function with period $P$ is also periodic with period $kP$, for any positive integer $k$. Given a certain time interval, a function can be periodic in just one sub-interval or can be periodic in several disjoint sub-intervals with different periods. A time interval when the function is periodic is a \emph{periodic interval}. As an example, the function of Fig.~\ref{fig:first-example} is periodic in two periodic intervals, with two different periodic patterns.

\subsection{The Role of the Metric}

Periodicity has been deeply studied in the case when the co-domain of $f$ has a metric (e.g.\ it is a subset of the real or of the integer numbers). Examples of this case that are in some way related to our problem follow.

In~\cite{vlachos2005periodicity} a hybrid technique is presented based on the combined usage of discrete Fourier transform and autocorrelation. The authors argue that autocorrelation techniques are useful for exploring large periods but do not allow to determine the period itself, while periodograms allow to use thresholds for noise filtering but are not accurate for short periods. Hence, they combine the two techniques.

A classification of Dance Music based on periodicity patterns is presented in~\cite{dixon2003classification}. Two methods are compared. The first performs onset detection and clustering of inter-onset intervals. The second uses autocorrelation. The autocorrelation-based approach gave better results.

Periodicity detection of local motion is studied in~\cite{tong2005periodicity}. The authors  proposed an approach for local motion analysis via periodicity detection under complex conditions. The work is mostly motivated by the difficulties of analyzing
local motion in the presence of clutters consisting of global and multi-object motions. The contribution is in the combination of the layered motion analysis and the autocorrelation of motion energy to estimate the basic motion periodicity. Interestingly, the field of interest requires dealing with a large amount of noise.

A classical method for investigating periodicities in disturbed series, with special reference to Wolfer's \emph{sunspot numbers} is presented in~\cite{yule1927method}. Sunspot numbers are analogous to the data that would be given by observations of a disturbed periodic movement, such as that of a \emph{pendulum} subject to successive small impulses. The method is effective especially for short periods and for short periodicities.

However, all the above techniques can only be used when the co-domain of $f(x)$ has a metric. In the case of Internet topology measures, $f$ is just a time-series of paths and there is no total order between them. Also, the domain is the discrete time since measures are usually performed at specific instants (e.g.\ each minute, each hour, etc.).

\subsection{Periodicity in Internet}

Some papers on periodicity issues have been presented also in the Internet research community. Even in this case we discuss those that are somehow related to our work.

Periodicity classification of HTTP traffic, to detect HTTP botnets, is discussed in~\cite{eslahi2015periodicity}. The HTTP botnets periodically connect to particular Web pages or URLs to get commands and updates from a botmaster. This identifiable periodic connection pattern has been used in several studies as a feature to detect HTTP botnets. The authors propose three metrics to be used in identifying the types of communication patterns according to their periodicity. Test results show that in addition to detecting HTTP botnet communication patterns with  accuracy, the proposed method is able to efficiently classify communication patterns into several periodicity categories.

Inferring the periodicity in large-scale Internet measurements is the purpose of~\cite{argon2013inferring}. The authors present two methods for assessing the periodicity of network events and inferring their periodical patterns. The first method uses power spectral density for inferring a single dominant period that exists in a signal representing the sampling process. This method is highly robust to noise, but is most useful for single-period processes. Hence, they present a method for detecting multiple periods of a single process, using iterative relaxation of the time-domain autocorrelation function. They evaluate these methods using extensive simulations, and show their applicability on real Internet measurements of end-host availability and IP address alternations.


Internet routing instability is studied in~\cite{labovitz1997internet}. The paper examines the network inter-domain routing information exchanged between backbone service providers at the major U.S. public Internet exchange points. The authors describe several unexpected trends in routing instability, and examine a number of anomalies and pathologies observed in the exchange of inter-domain routing information. They also show that instability exhibits strong temporal properties. They describe a strong correlation between the level of routing activity and network usage. The magnitude of routing information exhibits the same significant weekly, daily and holiday cycles as network usage and congestion. They essentially apply existing techniques to the time-series of the number of BGP updates.

Finally, ping roundtrip-time periodicity is extensively studied in~\cite{csabai19941} (based on the power spectrum) and Internet traffic periodicity is studied in~\cite{squillante2000internet}.

However, also in all the above cases the adopted techniques can only be used when the co-domain of $f(x)$ has a metric. To find some contributions where the co-domain of $f(x)$ does not have a metric we have to explore other research fields.

\subsection{Without a Metric}

DNA periodicity is studied in~\cite{silverman1986measure}. A Fourier transform of a sequence of bases along a given stretch of DNA is defined. The transform is invariant to the labelling of the bases and can therefore be used as a measure of periodicity for segments of DNA with differing base content. It can also be conveniently used to search for base periodicities within large DNA data bases. Unfortunately, the function can have at most four values and this makes this technique unuseful for our purposes.

A periodicity detection algorithm for databases of different types of data (e.g.\  automotive, video surveillance, and geography) is proposed in~\cite{parthasarathy2006robust}. The algorithm combines information from the time-frequency domain and from the autocorrelation space to find meaningful periods. If the fundamental period of a signal is $T$, in some cases the algorithm outputs $2T$. In such cases, the authors manually divide the periodicity by $2$ after a visual inspection of the signal. This manual inspection step makes the algorithm unlikely to scale and unsuitable for an on-line service.

Periodicity detection in time series databases is also studied in~\cite{elfeky2005periodicity} The authors define two types of periodicities. Whereas
\emph{symbol periodicity} addresses the periodicity of single symbols in the time series, \emph{segment periodicity} addresses the periodicity of portions of the series. Their proposed algorithm for segment periodicity detection uses the convolution in order to shift and compare the time series for all possible values of the period. However, the algorithm does not allow to automatically characterize the periodicity in terms of involved pattern and periodic interval.

\section{A Data Set of Traceroutes}\label{sec:data-set}

In recent years, measuring network performance and availability assumed a key operational role. This drastically increased the number of available Internet measurement platforms. Some of the open ones have distinguished features of accuracy and coverage.

RIPE Atlas is a platform with more than $9,000$ points of view consisting of small hardware devices (called \emph{probes}) able to perform various types of measurement, including traceroutes, against other hardware devices (called \emph{anchors}). These devices are distributed around the world, and probes are mostly at end-user connections~\cite{atlas}.

In our experiments we consider the IPv4 traceroutes performed by $9,738$ RIPE Atlas probes towards $258$ anchors in the week from May 1st to May 7th 2017. The total amount of probe-anchor pairs is $101,715$ since not all the possible pairs are active. Traceroutes are performed every $15$ minutes. The result of each traceroute is a path of IP addresses. In composing the paths we consider only the answer to the first of the three traceroute packets sent at each traceroute round. All these data are publicly available.

Observe that RIPE Atlas traceroutes are subject to IP address aliasing (see e.g.~\cite{Gunes:2009:RIA:1721711.1721715}), since no anti-aliasing preprocessing is performed by RIPE NCC before providing the traceroutes to the users of the service. Since our goal is to study the phenomenon on an as-is basis we did not perform any anti aliasing change in the data set.

\begin{figure*}[!htb]
\centering
   \begin{subfigure}[b]{0.75\textwidth}
     \includegraphics[width=1\linewidth]{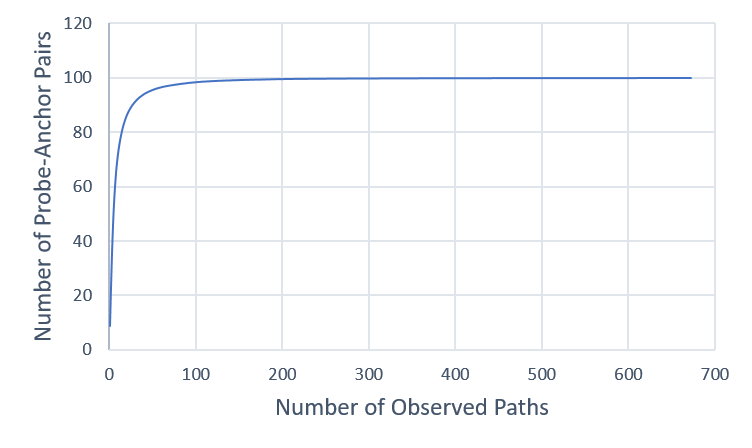}
     \caption{}
     \label{fig::pathDistribution} 
   \end{subfigure}

\begin{subfigure}[b]{0.75\textwidth}
   \includegraphics[width=1\linewidth]{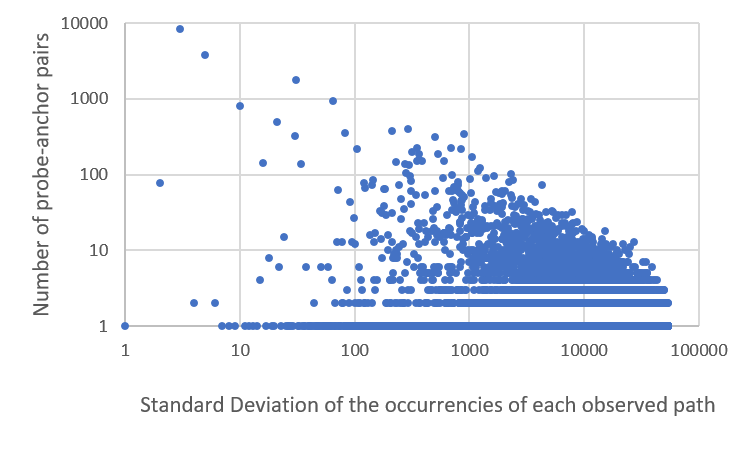}
   \caption{}
   \label{fig:standardDeviation}
\end{subfigure}
\caption[Basic features]{Basic features of the data set. (a) Distribution of the number of distinct paths with respect to the probe-anchor pairs. (b) Distribution of the number of probe-anchor pairs with respect to the standard deviation of the number of occurrences per distinct path.}
\label{fig:basic-features}
\end{figure*}

The purpose of Fig.~\ref{fig:basic-features}.a is to show how many distinct paths are observed by each probe-anchor pair. Observe that most of the pairs (roughly $85 \%$) observe less than $20$ paths. However, a few of them observe many more paths.

We may expect that a probe-anchor pair whose traceroutes are periodic have that each path is observed roughly the same number of times. As an example, this may happen for simple periodicities where the periodic pattern does not contain repeated paths.

Hence, we computed the number of occurrences of each distinct observed path for each pair. Fig.~\ref{fig:basic-features}.b shows the distribution of the number of probe-anchor pairs with respect to the standard deviation of the number of occurrences of each distinct path. The figure shows that real data are very complex and that it is hard to classify probe-anchor pairs according to the standard deviation of the number of observed paths.

\section{A Methodology for Detecting Periodicities}\label{sec:methodology}

In this section we describe an algorithm, composed of four steps, for detecting and characterizing periodicities. Its input is a time-series $x(t)$ of paths resulting from a sequence of traceroutes between a specific source-destination pair, where $x(t)$ is the path measured at time $t$. Its output is, possibly, a set of periodicities observed in $x(t)$ each consisting of a period, of a periodic pattern, and of a periodic interval with its starting/ending time. The algorithm can be tuned using a tolerance parameter $t_i$.

\subsection{Autocorrelation}

In the \emph{Autocorrelation Step} we compute a variation $R_{xx}(l)$ of the well known autocorrelation function of $x(t)$ as follows.

$$R_{xx}(l) = \sum_{n \in Z} x(n) \cdot x(n+l)$$

The \emph{dot} operator is a path matching operator that outputs one if $x(n) = x(n+l)$ and zero otherwise. When comparing paths containing asterisk we do not do any assumption on the missing IP addresses and consider asterisks as any other symbol in the path. Observe that an asterisk may not correspond to a routing change (for example, a router might reply with an ICMP packet only to every other packet with expired TTL).

\begin{figure*}[tb]
\centering
\fbox{\includegraphics[width=\linewidth]{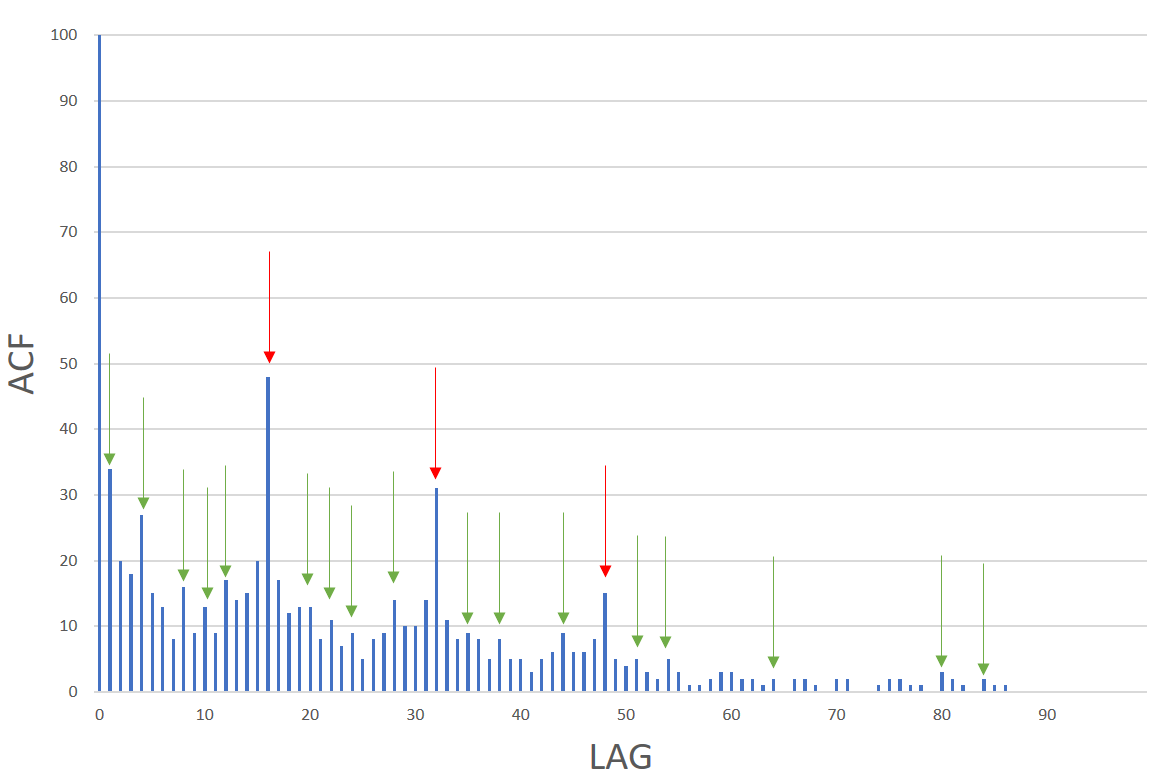}}
\caption{Autocorrelation (ACF) performed on the time-series of Fig.~\ref{fig:first-example}.}
\label{fig:algorithm}
\end{figure*}

The variable $l$ is called \emph{Lag}. The result of the autocorrelation performed on the time-series of Fig.~\ref{fig:first-example} is shown in Fig.~\ref{fig:algorithm}. In the figure each unit of lag corresponds to a time unit of 15 minutes, the default timing in Atlas anchoring measurements.

\subsection{Peaks-Detection}

In the \emph{Peaks-Detection} step we look for ``peaks'' in the $R_{xx}(l)$ function. Intuitively, peaks are clues of periodicity since they correspond to a large number of evenly spaced identical paths. To do that we simply look for local maxima. If no peak is found we conclude that $x(t)$ has no periodicities. Otherwise, we determine a set of peaks, each identified by a value $l$ of Lag.

If several peaks have been found it is possible that several periodicities are present. Also, different peaks may represent the same periodicity detected with overlapping periods. The arrows in Fig.~\ref{fig:algorithm} put in evidence twenty peaks.

\subsection{Peak-Clustering}

In the \emph{Peak-Clustering} step, peaks that are geometrically near each other are grouped and the resulting clusters are analyzed. Intuitively, we have that peaks with similar $l$ and similar $y$ are likely to correspond to the same periodicity.  In Fig.~\ref{fig:algorithm} the peaks with a red arrow are grouped (in this case the time between two consecutive of them is $240$ minutes) in the same cluster.

Peaks in the same cluster are then analyzed as follows. First, we order them according to their $l$. Second, we compute the $l$-distance (horizontal distance) between consecutive peaks. Third, we check if such distances are roughly the same or if they can become the same discarding one or more peaks (outliers). The peaks with a red arrow of Fig.~\ref{fig:algorithm} are roughly equi-spaced.

Clusters where distances are regular may correspond to periodicities and their inter-peak distances maybe their periods. Such clusters are the \emph{potential periodicities}. We have that at the end of this step Fig.~\ref{fig:algorithm} shows two clusters (red and green) that are both potential periodicities.

\subsection{Periodicity Characterization}

In the \emph{Periodicity Characterization} step for each potential periodicity we first verify if it can be considered a true periodicity. To do that $x(t)$ is split into sub-sequences with length corresponding to the potential period. Then consecutive subsequences are matched one against the other and are considered \emph{compatible} with the periodicity if they have a Hamming distance less or equal than tolerance $t_i$. If no compatible pairs of subsequences are found we conclude that $x(t)$ has no periodicities. Otherwise, we output a periodicity, the maximal subsequences of compatible intervals are glued into its periodic interval, its period is the length of the subsequence, and its pattern is the one with the highest number of repetitions among the glued intervals. 

In the example of Fig.~\ref{fig:algorithm} the cluster with green peaks does not pass this step and is hence discarded. The cluster with red peaks passes this step and we conclude that it corresponds to a true periodicity.

\section{Checking the Methodology In-Vitro}\label{sec:vitro}

In order to test the effectiveness of the algorithm of Section~\ref{sec:methodology} we randomly generated $5,000$ time-series of traceroutes and tested the algorithm against such time-series. 

Each of them consists of $10,000$ paths corresponding to one week of traceroutes performed every $60$ seconds.

\subsection{Generating the Time-Series}

\begin{figure*}[tb]
\centering
\fbox{\includegraphics[width=0.7\linewidth]{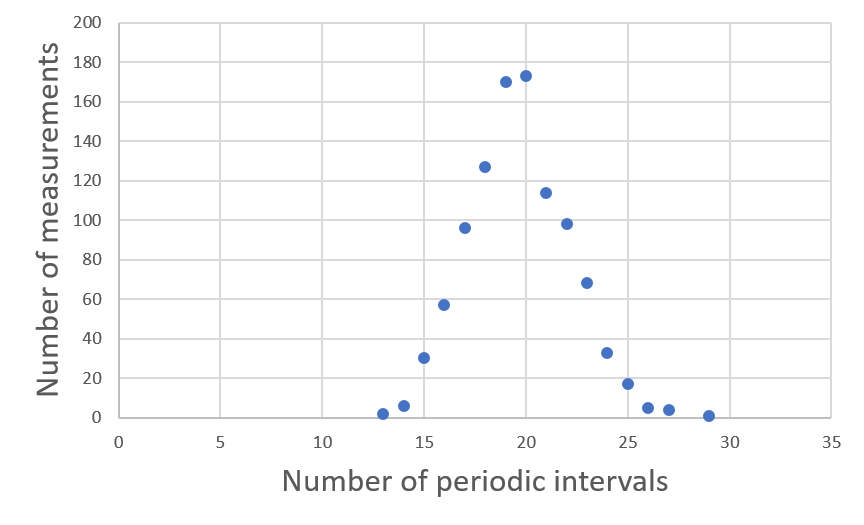}}
\caption{Relationship between time-series and periodicities.}
\label{fig:number-of-periodicities}
\end{figure*}

Each time-series is generated as follows. We first generate a periodic interval starting at time $t=0$. It consists of a random period in the range $\left[ 2, 30 \right]$ and a random periodicity pattern obtained by randomly selecting, for each instant of the period, a specific path in a group of pre-defined $30$ paths. Such parameters have been selected according to the analysis performed in Section~\ref{sec:data-set}. E.g., the choice to insert $30$ pre-defined paths into an observation is due to the analysis performed on the path distribution. In Section~\ref{sec:data-set} we showed that the most of the pairs (roughly $85\%$) observes less than 20 paths (see Fig.~\ref{fig:basic-features}.a). As consequence, the usage of $30$ distinct paths allows to work with a representative set of elements. The periodic interval is then obtained by randomizing the number of times the period is repeated.

Suppose that the periodicity ends at time $t_1$. We then generate a non-periodic random time interval from $t_1$ to $t_2$. The length of $\left[t_1,t_2\right]$ is randomized in such a way that it has the same probability distribution of the previous random periodic interval. The paths in the non-periodic interval are each randomized in the same set of paths used for periodic patterns. Observe that in most of the time-series of the cited public databases the periodic interval are quite often separated by intervals where very few paths (frequently just one) are found. Instead, we decided to fill non-periodic intervals with purely random paths in order to stress the algorithm.

We then randomize a new periodic interval starting from $t_2$ and keep on executing the same algorithm, alternating periodic and non-periodic intervals, until the end of the time-series is reached.

In order to have an even distribution of the length of the periodic patterns in the time-series, when the generation of the time-series is finished we perform a final randomization step re-shuffling the position of periodic and non-periodic patterns. Fig.~\ref{fig:number-of-periodicities} shows one of the features of the randomized data set. Namely, it shows  the number of time-series containing a given number of periodicities.

\subsection{Periodicities in the Random Time-Series}

We performed our experiments looking for periodicities in the generated time-series that contains $19,678$ periodicities. We have that $85.11 \%$ of the periodicities were found, with $14.89 \%$ of false negatives. On the other hand the algorithm detected $194$ periodicities that were not in the data set, with $0,9 \%$ of false positives. 

\begin{figure}[tb]
\centering
\fbox{\includegraphics[width=\linewidth]{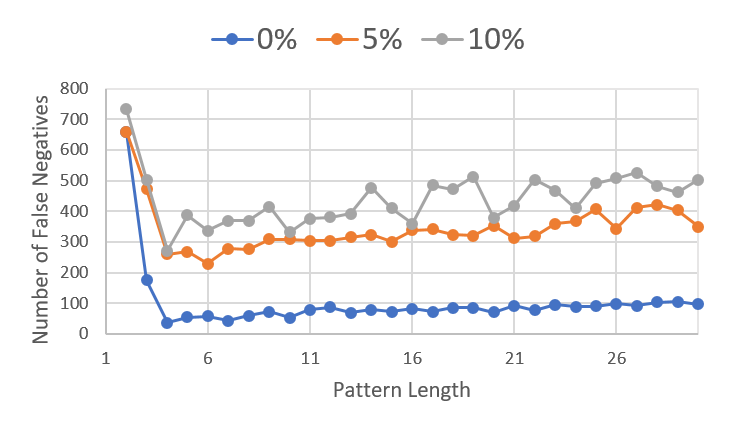}}
\caption{Distribution of the false negatives with respect to the length of the pattern.}
\label{fig:false-negative-patterns}
\end{figure}

\begin{figure}[tb]
\centering
\fbox{\includegraphics[width=\linewidth]{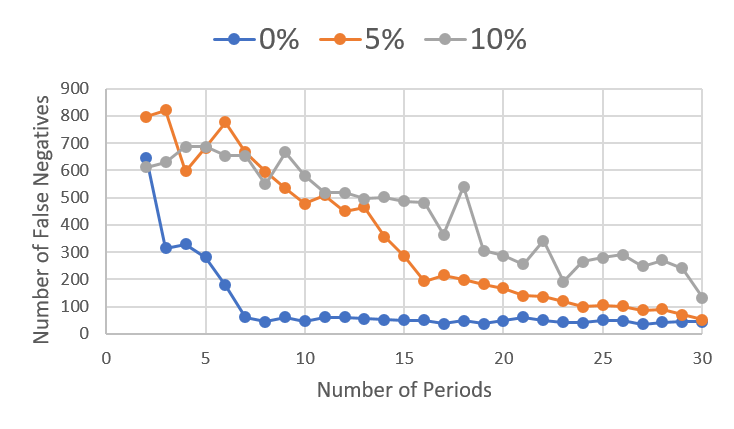}}
\caption{Distribution of the false negatives with respect to the number of periods}
\label{fig:false-negatives-periods}
\end{figure}

Also, among the periodicities that were correctly detected the algorithm was able to give a correct characterization for $99,2 \%$ of them. Intuitively, long periodic intervals and long periods are easier to detect with respect to short ones. Fig.~\ref{fig:false-negative-patterns} (blue curve) shows how false negatives are distributed with respect to the length of the period while Fig.~\ref{fig:false-negatives-periods} (blue curve) shows how false negatives are distributed with respect to the number of periods in the periodic interval. The figure shows that the algorithm is quite conservative in looking for periodicities. Observe that the conservative approach is largely justified by the goals pursued in this analysis. Our goal is the discovery of a phenomenon, so its possible underestimation does not affect our conclusions.

\subsection{Inserting Noise}

After the activities described above, we repeated the experiments inserting increasing percentages of noise in the time-series. 

We insert noise as follows. When an element of noise is inserted we select a random time $t_r$ in one of the periodic periods of the series and randomize with equal probability an action among those of the following set. Either we insert a random path at $t_r$, or we remove the path that is found at $t_r$, or we substitute the path at $t_r$ with a new random path. The noise is inserted only in the periodic intervals since it would be pointless to change the random paths of the non-periodic intervals into other random paths.

\begin{figure*}[tb]
\centering
\fbox{\includegraphics[width=\linewidth]{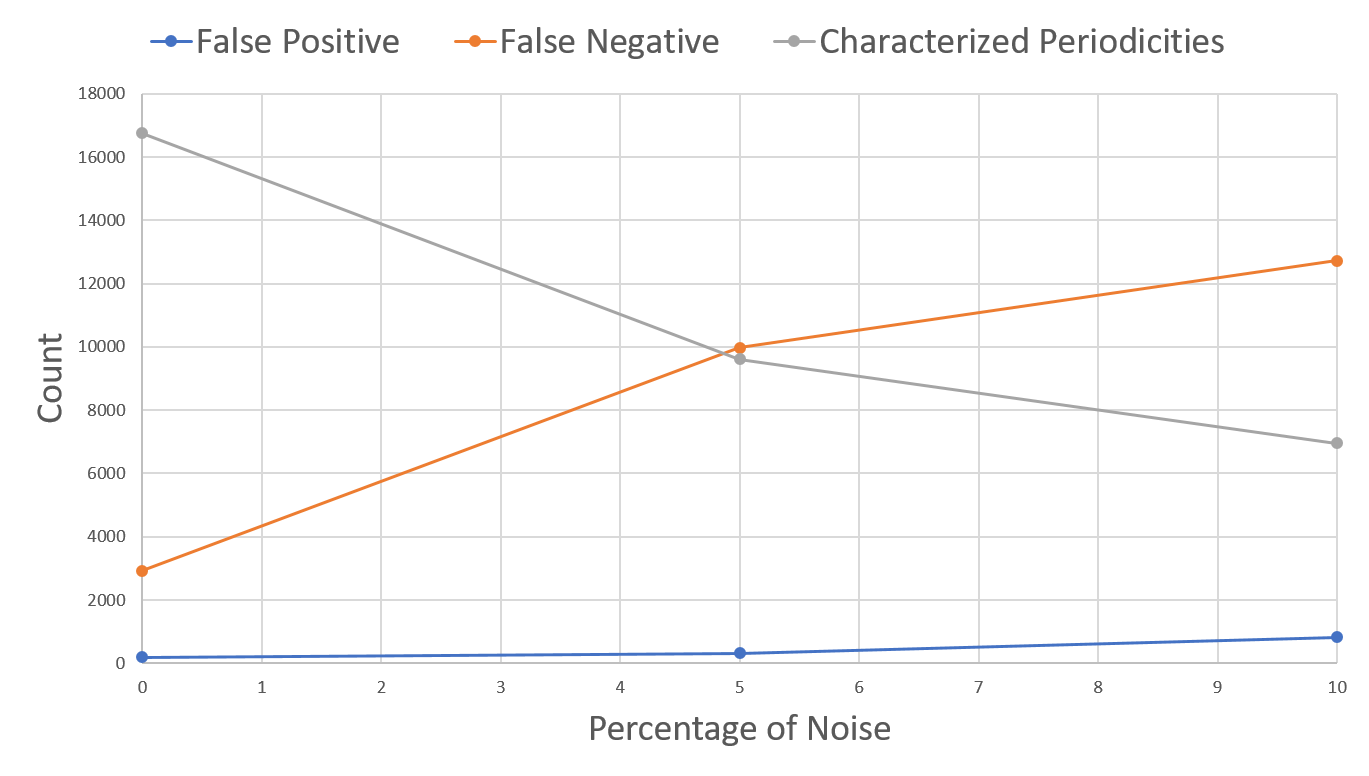}}
\caption{Performance of the algorithm in presence of noise. Percentage of false-negatives, of false-positives, and of correctly characterized periodicities. Recall that the total number of periodicities is $19,678$.}
\label{fig:noise}
\end{figure*}

In Fig.~\ref{fig:noise} we show how the algorithm presented in Section~\ref{sec:methodology} behaves in presence of noise. The \emph{percentage of noise} is the number of inserted elements of noise with respect to the number of paths belonging to periodic sequences.

The experiments performed in-vitro allowed us to determine a suitable value for the tolerance parameter $t_i$ of the algorithm, that was selected in order to have a very small number of false positives. Namely, if the the pattern contains less than $5$ paths we set $t_i=1$, otherwise we set $t_i$ to $10 \%$ of the pattern length. Hence, all the experiments reported in this paper have been done with this setting.

\section{Periodicity of Traceroutes of Public Data Sources}\label{sec:traceroute-periodicity}\label{sec:traceroute-experiments}

We applied the algorithm of Section~\ref{sec:methodology} to the data set of traceroutes described in Section~\ref{sec:data-set} with the purpose to check if periodic time-series of traceroutes exist in such data sources and, if yes, to determine their frequency.

\subsection{Periodicity in RIPE Atlas Traceroute Paths}

\begin{figure}[!htb]
\centering
   \begin{subfigure}[b]{0.53\textwidth}
     \includegraphics[width=0.9\linewidth]{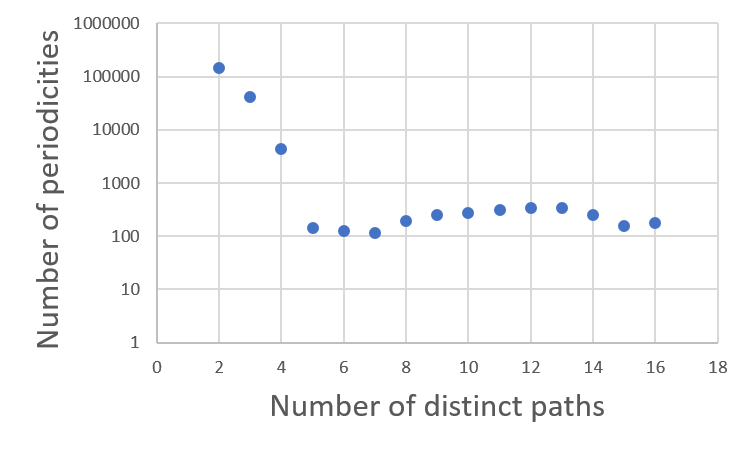}
     \caption{Distribution of the number of distinct paths\\ contained in periodicities.}
     \label{fig::data-set-periodicity-no-paris-60-seconds} 
   \end{subfigure}

\begin{subfigure}[b]{0.53\textwidth}
   \includegraphics[width=0.9\linewidth]{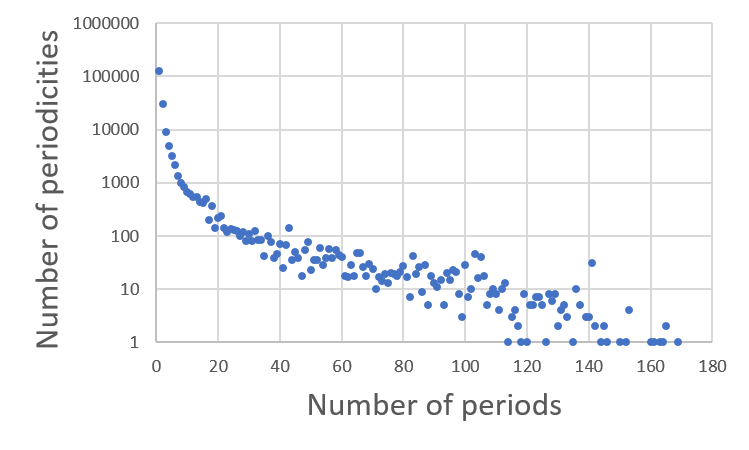}
   \caption{Distribution of the number of periods contained\\ in the periodic intervals.}
\end{subfigure}

\begin{subfigure}[b]{0.53\textwidth}
   \includegraphics[width=0.9\linewidth]{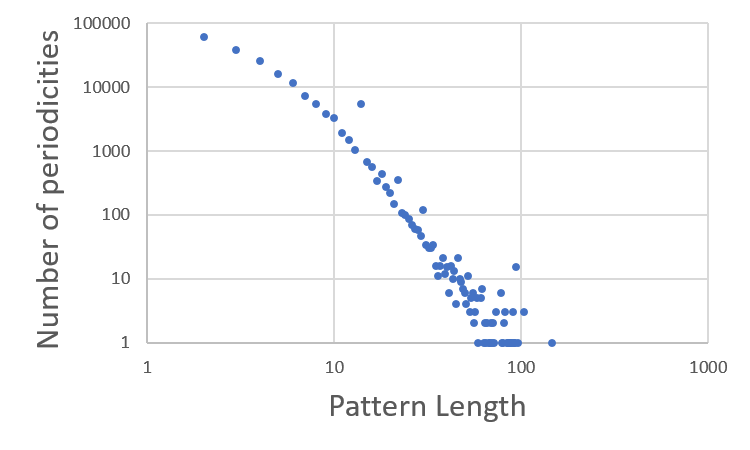}
   \caption{Distribution of the lengths of the patterns.}
\end{subfigure}

\begin{subfigure}[tb]{0.53\textwidth}
   \includegraphics[width=0.9\linewidth]{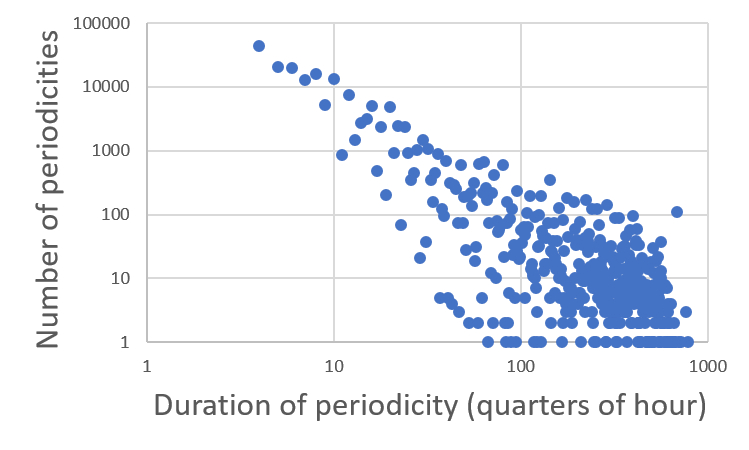}
   \caption{Distribution of the  durations of the periodicities.}
\end{subfigure}

\caption{Periodicities found in the data set.}
\label{fig:data-set-periodicity}
\end{figure}

We found that $36.02 \%$ of probe-anchor pairs have at least one periodicity in the time interval. Also, we have found a total of $186,403$ periodicities. Fig.~\ref{fig:data-set-periodicity} gives more details on the results of the experiment. Fig.~\ref{fig:data-set-periodicity}.a shows that about $122,000$ periodicities have a pattern that is composed of exactly two paths. Looking at such paths we have that about $60,000$ of them have an alternation of a path where all IPv4 addresses are present and a path where one of such addresses is substituted by an asterisk. Also, $180$ periodicities have a pattern of $16$ paths. Fig.~\ref{fig:data-set-periodicity}.b shows that about $148,000$ periodicities have a number periods repeated into the periodic interval that is less or equal than $10$. Fig.~\ref{fig:data-set-periodicity}.c shows that most of the periodicities (about $165,000$) have a periodic pattern will at most $10$ paths. Fig.~\ref{fig:data-set-periodicity}.d shows that the relationship between number of periodicities and duration is quite scattered. Most of them (about $131,000$) last less than two hours.

\subsection{Paris Traceroute and Plain Traceroute}

Up to now we did not consider the causes of the periodicities, since they are out of the scope of this paper. However, it is important to observe that traceroutes can be performed with different methods. Namely, the probes may either use Paris traceroute or traditional traceroute~\cite{paris}. In Paris traceroute packets are forged using different Paris-ids, whose value rotates in a range that varies from one to a maximum value that can be set by the user. The data set of Section~\ref{sec:data-set} has been computed using Paris traceroutes with maximum value of paris-id equal to $16$, that is the default for all RIPE-Atlas anchoring measurements.

\begin{figure*}[!htb]
\centering
   \begin{subfigure}[b]{\textwidth}
     \includegraphics[width=1\linewidth]{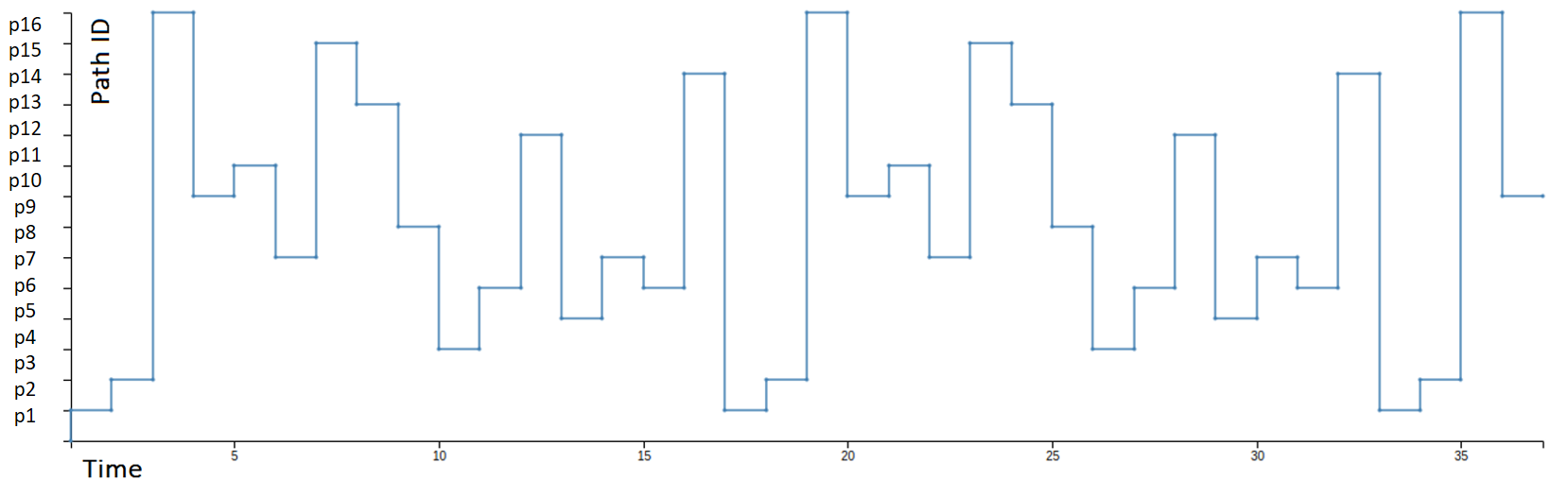}
     \caption{}
     \label{fig:parisBasse} 
   \end{subfigure}

\begin{subfigure}[b]{\textwidth}
   \includegraphics[width=1\linewidth]{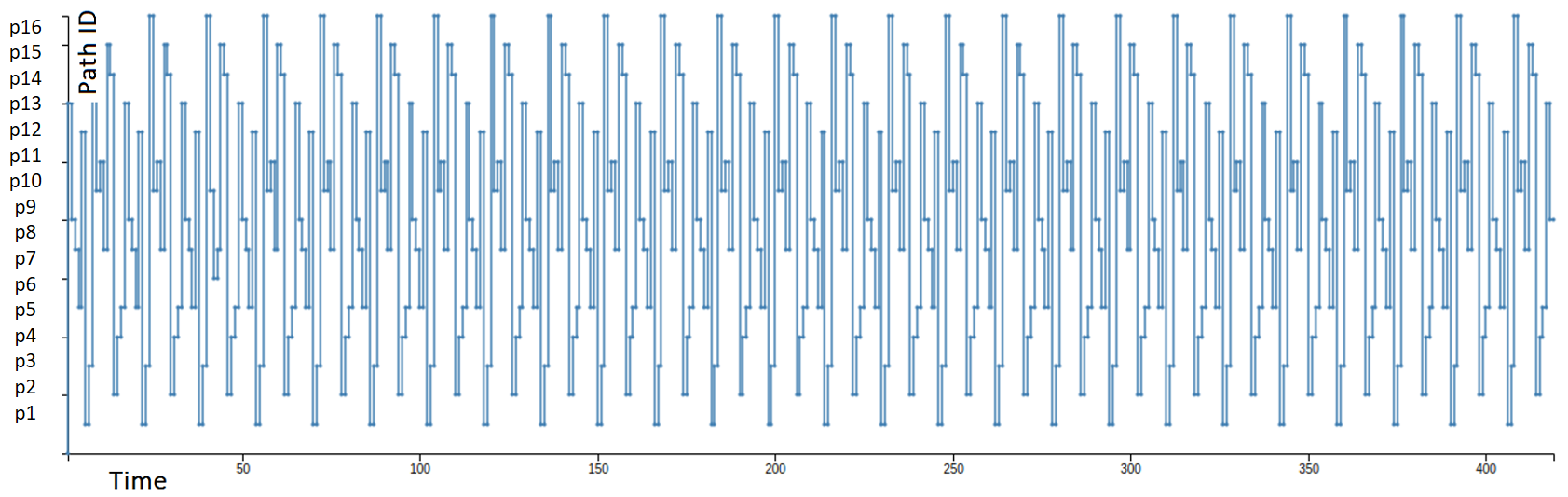}
   \caption{}
   \label{fig:parisAlte}
\end{subfigure}

\caption{(a) A periodicity found with one traceroute every $15$ minutes. (b) a periodicity found with one traceroute every minute. The two measures are done in the same interval of time and on the same probe-anchor pair.}
\label{fig:parisUsage}
\end{figure*}

The impact of using Paris traceroute on periodicity is made clear by the example of Fig.~\ref{fig:parisUsage}. Observe that exactly the same periodic pattern is found independently on the frequency of traceroutes. Most probably what happens is that packets forged by probes traverse different paths according to an hash function applied to the packet header~\cite{paris}. Also, since the periodic pattern of Fig.~\ref{fig:parisUsage} contains $16$ paths we may suppose the presence of a load-balancing components with at least $16$ forwarding options.

Once a periodicity is found, it is interesting to decide if Paris traceroute is, at least partially, responsible for it. To do that we can check if the periodic pattern contains a path that comes out in the periodicity in all cases when a specific Paris-id is used for the measure. In this case we argue that we are probably looking at a periodicity triggered by a load-balancer operating per-flow on Paris traceroute probes. In our data set $18.4 \%$ of the periodicities have this feature. Also, $72.6 \%$ of the corresponding patterns have less than $6$ paths.

The above discussion naturally opens the question whether periodicities can be found even if Paris traceroute is not used. Since measures without Paris traceroutes are not the default, we had to setup a special purpose set of measures on the probe-anchor pairs of our data set. To do that we had two types of problems: (i) performing ad-hoc traceroutes for all the pairs would be unfeasible and (ii) to have comparability of the results we had to perform ad-hoc measures in the same week of the data set. We solved the problem as follows. We performed a preliminary experiment in the week from April 24th to April 30th. We then spotted the periodic pairs of that week and randomized $60$ of them. Finally, in our reference week we performed the ad-hoc measures on those pairs, with the advantage to have two comparable data set in the same week.

\subsection{Changing the Frequency}

Another important issue to be discussed is about the possibility of finding periodicities changing the frequency of traceroutes. Hence, for the $60$ pairs described above we performed measures every $60$ seconds, that is the maximum frequency that is possible to set on RIPE Atlas probes.

\begin{figure}[!h]
\centering
   \begin{subfigure}[b]{0.49\textwidth}
     \includegraphics[width=0.9\linewidth]{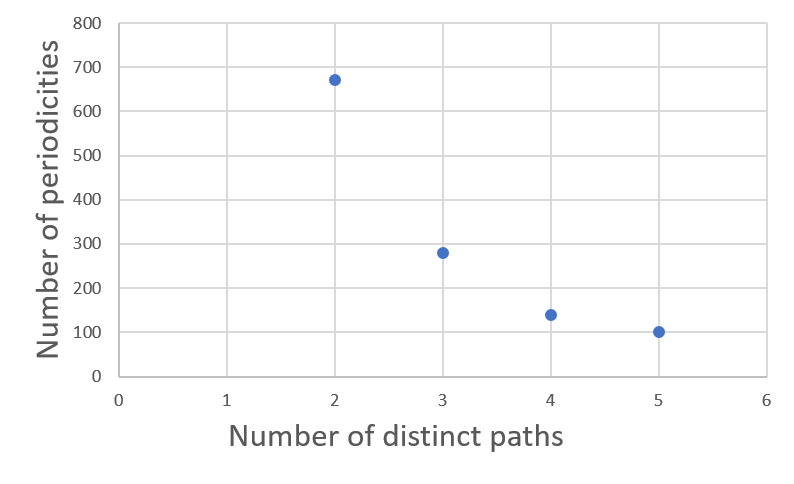}
     \caption{Distribution of the number of distinct paths contained in periodicities. Observe that the logarithmic scale is not used.}
   \end{subfigure}

\begin{subfigure}[b]{0.49\textwidth}
   \includegraphics[width=0.9\linewidth]{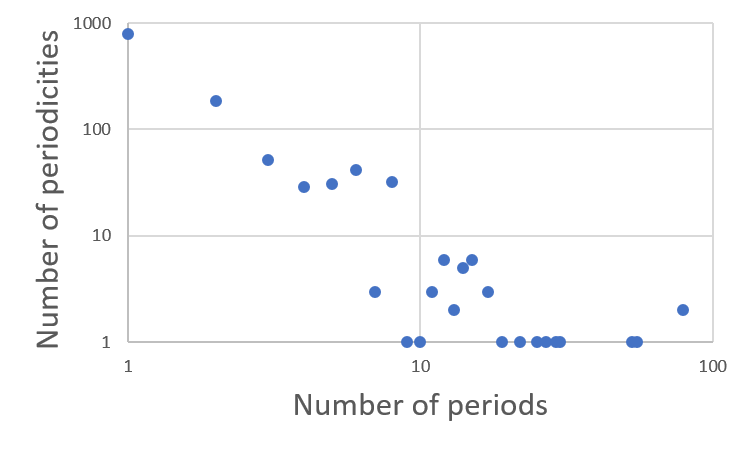}
   \caption{Distribution of the number of periods contained in the periodic intervals.}
\end{subfigure}

\begin{subfigure}[b]{0.49\textwidth}
   \includegraphics[width=0.9\linewidth]{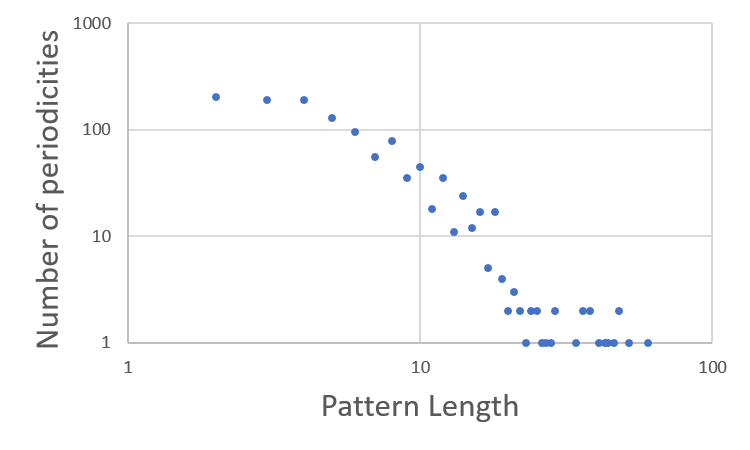}
   \caption{Distribution of the lengths of the patterns.}
\end{subfigure}

\begin{subfigure}[b]{0.49\textwidth}
   \includegraphics[width=0.9\linewidth]{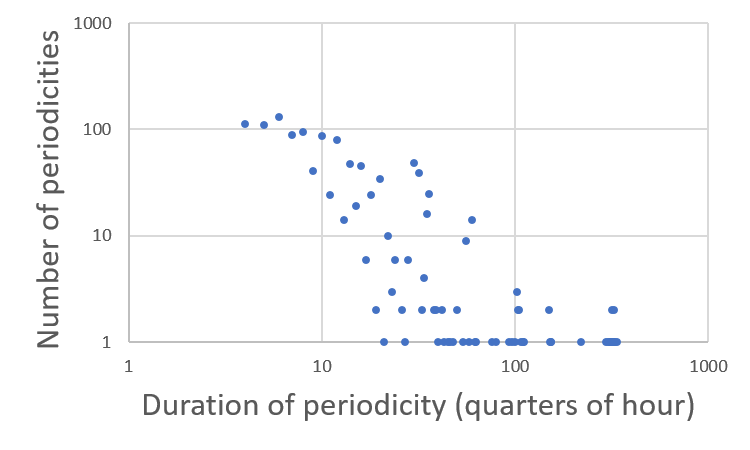}
   \caption{Distribution of the durations of the periodicities.}
\end{subfigure}

\caption{Periodicities found in the data set built not using Paris traceroute and with measures every $60$ seconds.}
\label{fig:data-set-periodicity-no-paris-60-seconds}
\end{figure}

We found that $36$ of the $60$ probe-anchor pairs with traceroutes performed every $60$ seconds have at least one periodicity in the time interval. Also, we have found a total of $1,194$ periodicities. Fig.~\ref{fig:data-set-periodicity-no-paris-60-seconds} illustrates the details of the experiment.

Fig.~\ref{fig:data-set-periodicity-no-paris-60-seconds}.a shows that $671$ periodicities have a pattern that is composed of exactly two paths. Looking at such paths we have that  $315$ of them have an alternation of a path where all IP addresses are present and a path where one of such addresses is substituted by an asterisk. Also, $101$ periodicities have a pattern of $5$ paths. Fig.~\ref{fig:data-set-periodicity-no-paris-60-seconds}.b shows that $1,158$ periodicities have a number periods repeated into the periodic interval that is less or equal than $10$. Fig.~\ref{fig:data-set-periodicity-no-paris-60-seconds}.c shows that most of the periodicities ($1,021$) have a periodic pattern will at most $10$ paths. Even in this case, Fig.~\ref{fig:data-set-periodicity-no-paris-60-seconds}.d shows that the relationship between number of periodicities and duration is quite scattered. Roughly half of them (about $534$) last less than two hours.

\begin{table*}[!htb]
  \centering
  \begin{tabular}{llcll}
   Prefix           & Origin AS                                               & Period & Start of Observation & End of Observation  \\ \cline{1-5}
110.170.17.0/24  & 134438 & 480       & 2017-05-14 15:15:33  & 2017-05-15 01:15:33 \\
93.181.192.0/19  & 13118  & 1800      & 2017-05-14 02:00:00  & 2017-04-14 12:00:00 \\
193.0.132.0/22   & 3203 & 710       & 2017-05-16 11:00:00  & 2017-05-16 21:00:00 \\
185.123.238.0/24 & 8296  & 1100      & 2017-05-15 4:00:00   & 2017-05-15 14:00:00 \\
196.250.233.0/24 & 37662  & 1200      & 2017-05-13 15:00:00  & 2017-05-14 1:00:00  \\
192.129.3.0/24   & 2614 & 450       & 2017-05-13 10:00:00  & 2017-05-13 20:00:00 \\
13.15.32.0/20    & 22390 & 820       & 2017-05-13 00:00:00  & 2017-05-13 10:00:00 \\
91.193.202.0/24  & 25211  & 420       & 2017-05-12 06:00:00  & 2017-05-12 16:00:00 \\
133.69.128.0/19  & 2523  & 450       & 2017-05-17 02:00:00  & 2017-05-17 12:00:00 \\
133.69.128.0/20  & 2523  & 350       & 2017-05-17 02:00:00  & 2017-05-17 12:00:00 \\
154.66.175.0/24  & 25543  & 620       & 2017-05-16 14:15:00  & 2017-05-17 00:15:00
  \end{tabular}
  \caption{Prefixes with a periodicity. The Period column shows the duration of the spotted period in seconds. The right columns specify the observed interval of time (roughly ten hours).}
  \label{tab:1}
\end{table*}

\section{Periodicity of BGP Updates}\label{sec:bgp-experiments}

We apply our algorithm to another important set of topology measures. Namely, we consider the BGP updates collected by the RIPE Routing Information Service (RIS)~\cite{riperis}. Since year 2001, the RIPE RIS collects and stores BGP updates from several locations around the world. The routers where such updates are gathered are called \emph{collector peers}. We consider $151$ of those collector peers.

While traceroute measures are active measures, since they inject packets into the network waiting for a reply, the BGP updates gathered by collector peers are passive measures, since they just collect information from the network without performing any action.

\subsection{State of the Internet}

Given an IPv4 prefix $\pi$, we are interested to spot periodicities related to $\pi$ and visible in the entire Internet. To do that we define the \emph{state} of the Internet with respect to $\pi$ as follows. We subdivide a time interval of interest in atomic instants each lasting one second (the maximum time granularity for BGP updates). The \emph{state} of a collector peer $c$ at time $t$ with respect to $\pi$ is the AS-path used by $c$ at time $t$ for reaching $\pi$. The \emph{state} of the Internet at time $t$ with respect to $\pi$ is the set of the states of all the collector peers. Roughly speaking, the state at time $t$ is how the (known part of) Internet reaches $\pi$ at time $t$. In this case the value of function $f$ at time $t$ is the state of Internet at time $t$.

We applied our algorithm to detect if the evolution of the state of the Internet has some periodicity. However, before doing that we needed to get further evidence of the effectiveness of the algorithm. In fact, the experiments performed in Section~\ref{sec:vitro} are all targeted to check the validity of the algorithm against time-series of paths obtained from traceroutes and hence it is not completely clear that those results apply even in the case where BGP updates are considered.

\subsection{Beacons}

In Internet there are prefixes that are announced and withdrawn on a regular basis, that are called \emph{beacons}~\cite{beaconsripe}. 

They are used for networking experiments and the sequence of announcements-withdrawals involving them is as follows. First, an announcement is issued, then after two hours it is withdrawn, then after two hours it is announced again. The origin router keeps on doing this ``forever'' with an overall period of four hours.

Hence, we tested the algorithm against all IPv4 beacons ($14$) listed in~\cite{beaconsripe} for $24$ hours and for each of them we were able to spot just one periodicity lasting four hours. For the case of BGP updates we slightly changed the algorithm. Namely, when the autocorrelation is performed the dot operator is redefined such that it outputs one if $95 \%$ the states of the Internet at time $t$ and at time $t+l$ coincide but for at most $5 \%$ of the collector peers. This is done to tolerate little temporal anomalies.

\subsection{Experiments with the Most Active Prefixes}

\begin{figure*}[tb]
\centering
\fbox{\includegraphics[width=\linewidth]{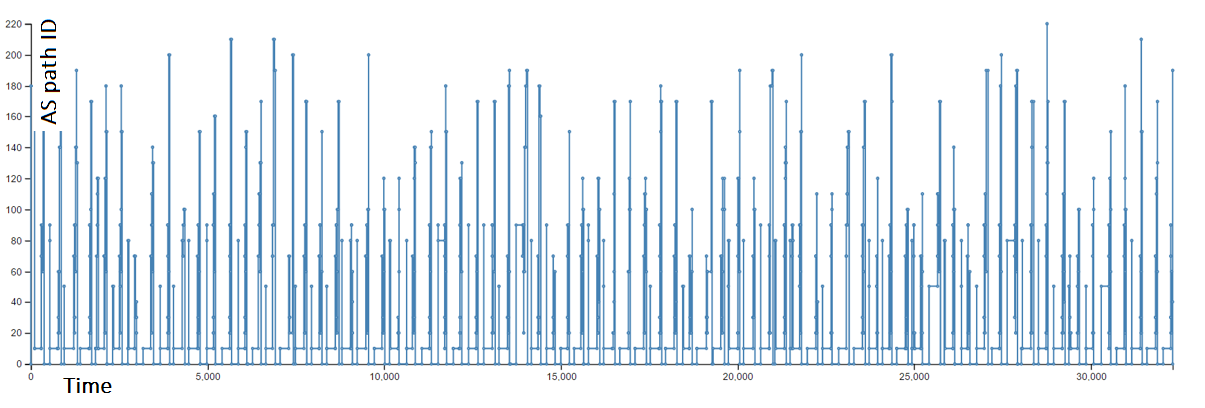}}
\caption{The time series of the states of  collector peer 01-195.66.226.20 with respect to prefix 66.19.194.0/24. Each value of the $y$-axes corresponds to a distinct AS-path.}
\label{fig:gdbASinverso}
\end{figure*}

\begin{figure}[tb]
\centering
\fbox{\includegraphics[width=\linewidth]{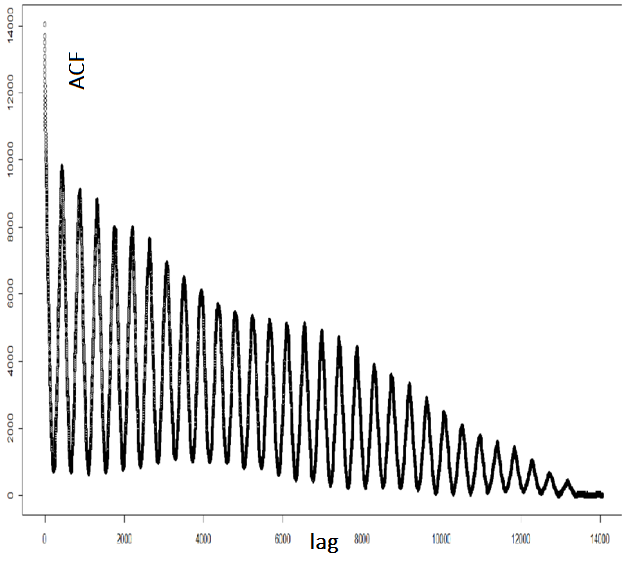}}
\caption{Output of the autocorrelation applied to the states sequence of collector peer 01-195.66.226.20.}
\label{fig:acfInverso}
\end{figure}

At this point we performed our experiments focusing on the $50$ most active prefixes of the Potaroo Web site~\cite{potaroo} in the week from May 7th to May 14th 2017. Such prefixes are involved in that week in more than $400,000$ updates, ranging from about $20,000$ for the most active to the about $4,000$ for the most quiet.

For each of the prefixes we extracted the corresponding updates from the RIPE RIS and considered only the ten hours with the highest number of updates. We obtained the results illustrated in Table~\ref{tab:1}. The table shows the $11$ of the $50$ prefixes where we have found   a periodicity.

\subsection{A Special Case}

We applied our algorithm to analyze the evolution of the state of the Internet also before the the week from May 7th to May 14th. 

\begin{figure*}[tb]
\centering
\fbox{\includegraphics[width=0.64\linewidth]{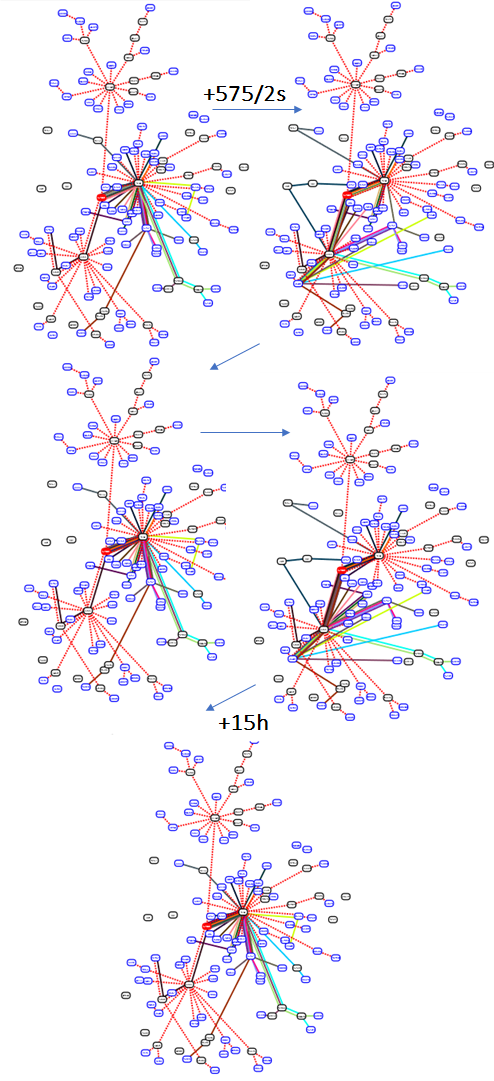}}
\caption{Evolution of the state of Internet with a period of 580 seconds shown with BGPlay visualization system. The states to the left are identical.}
\label{fig:bgplay}
\end{figure*}

A very clear example of periodicity that we have found is depicted by using the popular BGPlay visualization system~\cite{colitti2005visualizing} in Fig.~\ref{fig:bgplay}.  
With respect to prefix 45.42.41.0/24, the state shows a period of 580 seconds. Observe that the frames on left show almost completely the same configuration.

Looking at the details of the detected periodicities we observed that, between 02:00 AM and 11:00 AM of Apr 18 2017, with respect to prefix 110.170.10.0/24, the collector peer identified with 01-195.66.226.20 shows a period of 450 seconds. During the periodicity it alternates between AS-paths 56730-51945-2914-1299-7029-6316 and 56730-51945-1299-2914-23352-6316. Observe that the pair of adjacent ASes 1299 and 2914 appear alternatively in this order and in the opposite order. According to~\cite{cdrv-wrrirfspr-ton-11} this may correspond to the presence of a \emph{dispute reel} in the control plane of the routers traversed by the updates. As far as we know this is the first time that this configuration is observed to oscillate in the wild. Of course this may even correspond to a configuration that has been set by purpose. Fig.\ref{fig:gdbASinverso} and Fig.\ref{fig:acfInverso} show the sequence of AS paths and the output of the autocorrelation applied to the sequence, respectively.


\section{Conclusions}\label{sec:conclusions}

We presented a methodology for inferring periodic behavior in Internet topology measures. We believe that this can be a useful card to compose the puzzle of methods and tools that is needed for fully understanding the complex dynamics of such measures.

We have shown that finding Internet topology measures that exhibit a periodic behavior is frequent both for traceroutes and for BGP update collections.

The RIPE NCC is considering the possibility of putting at disposal of all the network operators a service, called \emph{Periodicity-as-a-Service}, for detecting the periodicities of RIPE Atlas traceroute measurements. The service has been announced and discussed with operators at RIPE-74. A prototype implementation that exploits the algorithm presented in this paper is available at http://atlas.ripe.net/periodicity. 

\clearpage


















\bibliographystyle{acm}
\bibliography{main} 




\end{document}